\documentclass[fleqn,10pt]{wlscirep}
\usepackage[utf8]{inputenc}
\usepackage[T1]{fontenc}
\usepackage{graphicx}
\usepackage{dcolumn}
\usepackage{bm}
\usepackage[utf8]{inputenc}
\usepackage{graphicx}
\usepackage{color}
\usepackage{amssymb}
\usepackage{subfigure}
\usepackage{epsfig}
\usepackage{float}
\usepackage{lipsum}
\usepackage{cases}
\usepackage{appendix}
\usepackage{cancel}
\usepackage{soul}
\usepackage{cite}

\newcommand{\D}{\boldsymbol{D}}

\newcommand{\G}{\boldsymbol{G}}

\newcommand{\F}{\boldsymbol{F}}
\newcommand{\M}{\mathrm{\mathbb{M}}}

\newcommand{\B}{\boldsymbol{B}}

\newcommand{\T}{\boldsymbol{\mathrm{T}}}
\newcommand{\A}{\boldsymbol{A}}
\newcommand{\1}{\mathbb{I}}

\newcommand{\CC}{\boldsymbol{C}}

\newcommand{\qq}{\boldsymbol{z}}
\newcommand{\J}{\boldsymbol{J}}

\newcommand{\dif}{\mathop{}\!\mathrm{d}}
\newcommand{\Det}{\mathop{}\!\mathrm{Det}}

\newcommand{\diag}{\mathop{}\!\boldsymbol{\mathrm{diag}}}

\newcommand{\rr}{\boldsymbol{r}}

\newcommand{\0}{\boldsymbol{0}}

\newcommand{\xxi}{\boldsymbol{\xi}}

\newcommand{\xchi}{\boldsymbol{\chi}}

\newcommand{\U}{\boldsymbol{U}}

\newcommand{\dbar}{\lower0.15ex\hbox{$\mathchar'26$}\mkern-12mu \dif}

\title{Brownian magneto-gyrator as a tunable microengine}

\author[1]{Iman Abdoli}
\author[2]{Ren\'e Wittmann}
\author[3]{Joseph Michael Brader}
\author[1,4]{Jens-Uwe Sommer}
\author[2]{Hartmut L\"owen}
\author[1,4,*]{Abhinav Sharma}
\affil[1]{Leibniz-Institut  f\"ur Polymerforschung Dresden, Institut Theorie der Polymere, Dresden, 01069, Germany}
\affil[2]{Institut f\"ur Theoretische Physik II, Weiche Materie, Heinrich-Heine-Universit\"at D\"usseldorf, D\"usseldorf, 40225, Germany}
\affil[3]{Department de Physique, Universit\'e de Fribourg, CH-1700 Fribourg, Switzerland}
\affil[4]{Technische Universit\"at Dresden, Institut f\"ur Theoretische Physik, Dresden, 01069,  Germany}
\affil[*]{sharma@ipfdd.de}

\begin{abstract}

A Brownian particle performs gyrating motion around a potential energy minimum when subjected to thermal noises from two different heat baths. Here, we propose a magneto-gyrator made of a single charged Brownian particle that is steered by an external magnetic field. Key properties, such as the direction of gyration, the torque exerted by the engine on the confining potential and the maximum power delivered by the microengine can be  tuned by varying the strength and direction of the applied magnetic field. Further tunability is obtained by rotating the potential in the plane perpendicular to the direction of the magnetic filed. 
We show that in this generic scenario, the microengine can be stalled and even reversed by the magnetic field. Finally, we highlight a property of the magneto-gyrator that has no counterpart in the overdamped approximation--the heat loss from the hot to cold bath requires explicit knowledge of the mass of the particle. Consequently, the efficiency of the microengine is mass-dependent even in the overdamped limit.

\end{abstract}
\begin{document}

\flushbottom
\maketitle

\thispagestyle{empty}

\section*{Introduction}

A heat engine converts thermal energy into mechanical energy while operating between hot and cold heat reservoirs. The working principles of a macroscopic heat engine are well-understood within the framework of thermodynamics~\cite{carnot1872reflexions}. Whereas fluctuations can be safely ignored in macroscopic systems, they are dominant in systems at nano- and micro-meter length scales \cite{bustamante2005nonequilibrium, jarzynski2011equalities, bo2013entropic, martinez2017colloidal}. Nevertheless, even on the nanoscale, work can be extracted, for instance, by rectification of thermal noise by Brownian motors~\cite{hanggi2009artificial}. Another example of a nanoscale device is the Brownian refrigerator in which a Brownian motor, driven by a temperature gradient,  performs cooling upon loading~\cite{van2006brownian}. Our current understanding of such nano devices is based on stochastic thermodynamics, a powerful theoretical framework that extends the concepts of heat, work and entropy production to small systems dominated by thermal fluctuations~\cite{seifert2005entropy,sekimoto2010stochastic}. Less than a decade ago, Blickle and Bechinger~\cite{blickle2012realization} devised the first experimental realisation of a microscopic heat engine. They  made a microscopic Stirling engine, where a single Brownian particle was subjected to a time-dependent optical trap and periodically coupled to different heat baths. Ever since, several other microscopic heat engines have been realised~\cite{quinto2014microscopic,martinez2016brownian, krishnamurthy2016micrometre,schmidt2018microscopic,ciliberto2017experiments}.

A Brownian gyrator is probably the simplest microscopic engine that can be conceived. As the name suggests, it constitutes a single Brownian particle, coupled to two thermostats, gyrating around a generic potential energy minimum~\cite{filliger2007brownian}. This exactly solvable model serves as a paradigmatic microscale heat engine that delivers systematic torque on the confining potential~\cite{filliger2007brownian,bo2013entropic, stark2014classical, verley2014unlikely, fogedby2017minimal,dotsenko2013two,mancois2018two, cerasoli2018asymmetry, murashita2016overdamped, holubec2017thermal, nascimento2020memory, nascimento2020stationary, chang2021autonomous}. The idea of Brownian gyration has been generalised to other systems such as resistor-capacitor circuits, subjected to different thermal noises, coupled by a third capacitor~\cite{chiang2017electrical,ciliberto2013heat} and colloidal systems~\cite{ciliberto2013heat, ghanta2017fluctuation, gonzalez2019experimental}. Experimentally, Brownian gyrator was realised recently by optically trapping a charged colloidal particle in an elliptical potential and coupling to two different thermostats along perpendicular directions using an electric noise~\cite{argun2017experimental}.  

\begin{figure}[t]
\centering
\includegraphics[width=0.5\linewidth]{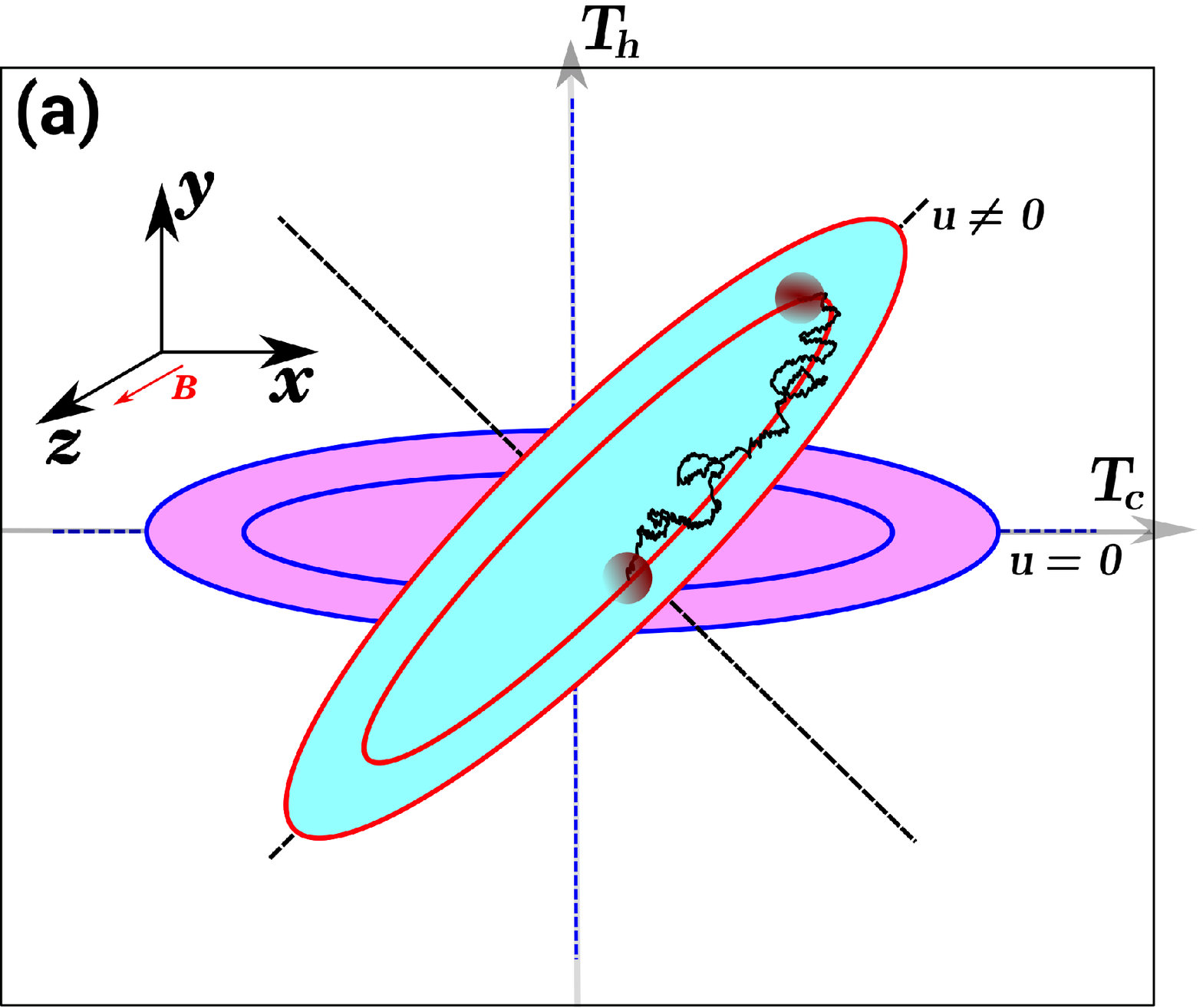}
\includegraphics[width=0.475\linewidth, height=7cm]{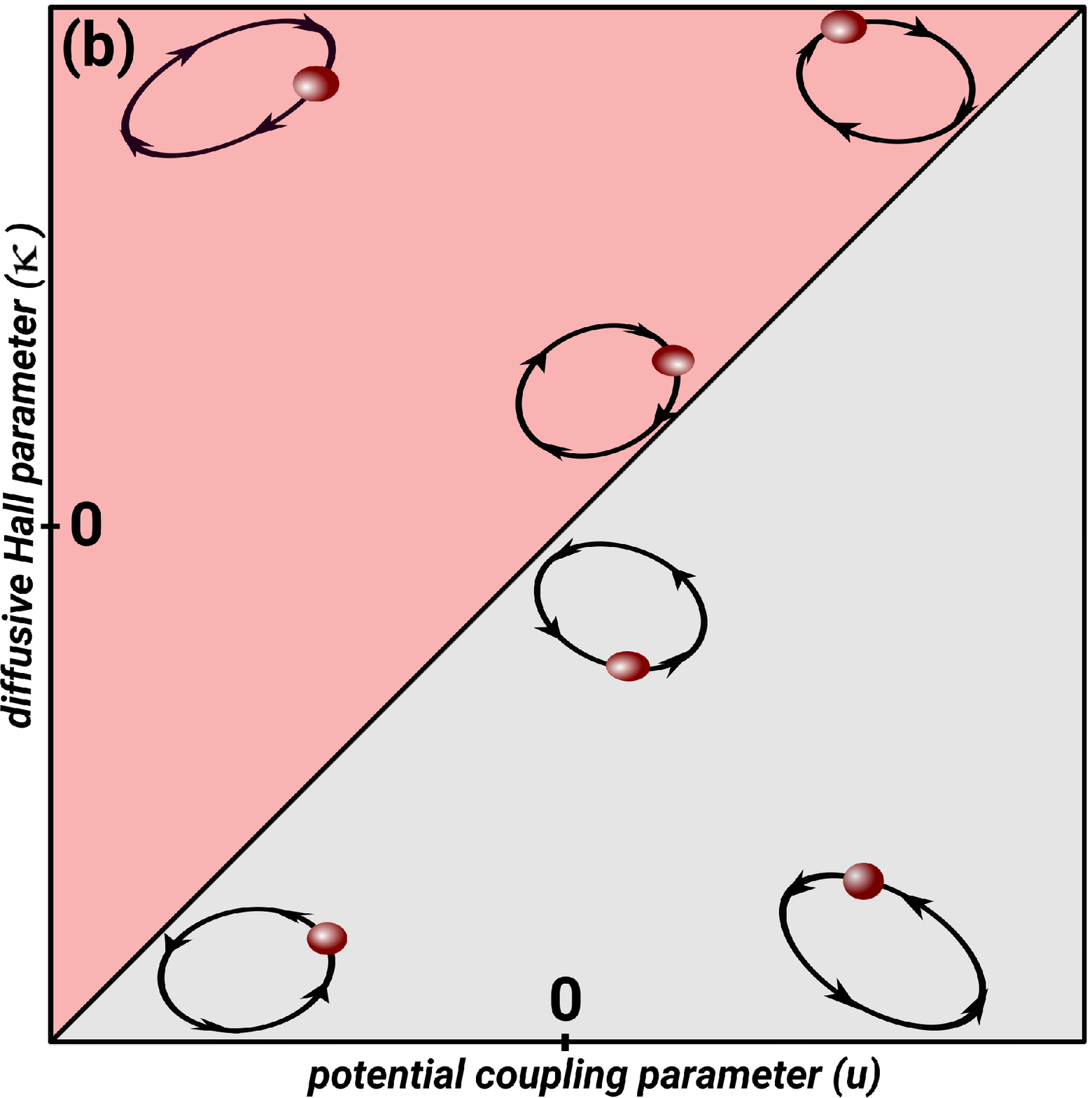}
\caption{  A single charged Brownian particle, steered by an external magnetic field $B$ and trapped in a rotationally asymmetric potential, performs gyration when subjected to different thermal noises from cold, $T_c$, and hot, $T_h$, heat baths coupled to its $x$ and $y$ degrees of freedom, respectively. Figure (a) schematically depicts the diffusion of the particle in a magnetic field in the $z$ direction and under the influence of the harmonic potential $V(x,y)=k[(x^2+\alpha y^2)/2+uxy]$ with the parameters $k$ and $\alpha$. Here $u$ is the potential coupling parameter which correlates the spatial degrees of freedom. The generic scenario is considered when the principal axes of the potential, shown by dashed lines, are misaligned with the temperature axes, namely if $u\neq 0$. Note that this condition (i.e., $u\neq 0$) is not necessary for the system to operate as an engine, yet is considered for further tunability of the microengine. (b) Schematic illustration of the average torque generated by the microengine for different values of the diffusive Hall parameter $\kappa=qB/\gamma$ and the potential coupling parameter $u$. The solid, oblique line represents zero torque in the system corresponding to stalled engine. The reddish region corresponds to the clockwise torque while the gray region depicts the counterclockwise torque. As can be seen, a torque can be generated even if the potential coupling parameter is zero upon applying an external magnetic field. The generated torque can be tuned by varying the magnitude and direction of the applied magnetic field: it changes the direction by reversing the direction of the applied magnetic field if $u=0$ or by additionally changing the sign of the potential coupling parameter when it is not zero. Note that latter can be also done by further tunability of the applied magnetic field for a fixed $u$.}

\label{fig1:schema}
\end{figure}

The simplicity of the Brownian gyrator is its most appealing feature. Nonetheless, it leaves little room for external manipulation of the engine's properties. Here, we propose a magneto-gyrator made of a single  Brownian particle, that retains the simplicity of a Brownian gyrator while being steered by an external field.  Our magneto-gyrator  is made of a single charged Brownian particle subjected to Lorentz force due to a constant external magnetic field. The particle is simultaneously subjected to different thermal noises along its spatial degrees of freedom and trapped in a rotationally asymmetric potential (see Fig.~\ref{fig1:schema}(a)). The fundamental properties of the gyrator, such as the direction of gyration, the torque exerted by the engine on the confining potential and the maximum power delivered by the microengine can be tuned by varying the strength and direction of the applied magnetic field. In contrast to previous studies, a magneto-gyrator does not require special alignment of the temperature axes with respect to the potential (See Fig.~\ref{fig1:schema})~\cite{filliger2007brownian,bo2013entropic, stark2014classical, verley2014unlikely, fogedby2017minimal,dotsenko2013two}. In fact rotating the potential with respect to the temperature axis provides another handle on tuning the magneto-gyrator. In this generic scenario, the microengine can be stalled and even reversed by tuning the applied magnetic field. 

Besides tunability, the study of a magnetic field in microengines is interesting from a fundamental theoretical perspective. In a pioneering work, Benenti and Casati showed that in systems with broken time-reversal symmetry, for instance via an external magnetic field~\cite{harman1963theory}, there is an additional freedom in the design of high performance microdevices than their time-reversal-symmetric counterparts.
Their study spurred tremendous research into work, heat, and efficiency of microengines in which the time-reversal symmetry is broken~\cite{brandner2013multi, brandner2015bound,yamamoto2016efficiency,polettini2017carnot, lee2017carnot, shiraishi2015attainability, koning2016engines}. 
Recently, the effect of Lorentz force has been studied in active and passive colloidal systems which are dominated by overdamped dynamics~\cite{jimenez2006brownian,filliger2007kramers, jimenez2008brownian, tothova2010hydrodynamic, lisy2013brownian, jimenez2013brownian,chun2018emergence, vuijk2019anomalous, vuijk2019lorenz, abdoli2021stochastic, matevosyan2021non}. 

The energy transfer in the microengine is not only mechanical, but also thermal in the form of heat. Calculating the rate of heat loss from the hot to the cold bath when the magneto-gyrator is operating as an engine, we study the efficiency of the magneto-gyrator. 
Surprisingly, we find that the heat loss in the magneto-gyrator requires explicit knowledge of the particle's mass~\cite{celani2012anomalous}. This finding leads to a paradoxical conclusion: even in the overdamped scenario where inertial effects have long decayed, one requires the knowledge of mass to calculate the heat losses in the system. The paradox is resolved by considering the motion of the particle in the underdamped regime. The particle picks up kinetic energy from the hot bath and then dissipates a part of it in the cold bath as it is curved by the applied magnetic field. The dissipation of energy from the hot to the cold bath occurs on the time scale of velocity correlations which scale inversely with the mass of the particle. The steady state is thus characterised by correlations between different velocity components~\cite{abdoli2020correlations}. This implies that even in the overdamped regime one cannot disregard the inertia-dependent heat loss in the engine. Using values from the literature~\cite{argun2017experimental, kahlert2012magnetizing}, we show that a magneto gyrator could be realised in dusty plasmas with an efficiency of $\sim 0.2\eta_{CA}$ at maximum power, where $\eta_{CA} $ is the Curzon-Ahlborn upper bound~\cite{curzon1975efficiency}.

The paper continues as follows: we first introduce the model of a Brownian magneto-gyrator in the overdamped regime for which we calculate average torque, work and heat loss. We then study the tunability of the microengine by varying the applied magnetic field. Finally, we answer the question of how efficient is the microengine.

\section*{Results} 
\subsection*{Brownian Magneto-Gyrator}
A Brownian magneto-gyrator is made of a single Brownian particle with charge $q$ steered by a constant magnetic field $B$ and trapped by the conservative force $\F_c=-\nabla V(\rr)$. We consider a magnetic field in the $\hat{z}$ direction such that $\B = B\hat{z}$ and hence the particle's motion along this direction is not affected by the applied magnetic field. Consequently, we effectively have a two-dimensional system with the particle's position $\rr=(x, y)^\top$ where $\top$ indicates the transpose.   
The thermal fluctuations of unequal strength, proportional to the cold and hot heat bath temperatures $T_c$ and $T_h$, are supplied along the two Cartesian coordinates $\hat{x}$ and $\hat{y}$, respectively. Here $V(\rr)=\frac{1}{2}\rr^\top\cdot\U\cdot \rr$ is the potential where the matrix $\U$ is given as
\begin{equation}
\label{matrixU}
\U = k\left( \begin{array}{cc}
1 & u \\
u & \alpha \\
\end{array}\right), 
\end{equation}
where $k$ is the stiffness of the potential, $\alpha$ is a dimensionless parameter quantifying the difference in the stiffness in the $x$ and $y$ directions, and $u$ is the potential coupling parameter which correlates the spatial degrees of freedom. Note that the stability condition implies that $|u|< \sqrt{\alpha}$. The eigenvectors of the matrix $\U$ correspond to the principal axes of the potential, which are shown by dashed lines in Fig.~\ref{fig1:schema} (a). This figure shows a schematic of the system where the alignment of the principal axes of the potential with the temperature axes is specified by the potential coupling parameter. In the case of $u=0$ the principal axes of the potential align with the temperature axes. The overdamped dynamics of the system can be described by the following Langevin equation (see Methods and Supplemental Material for details)
\begin{equation}
\label{langevinequaion:overdamped}
\dot{\rr}(t) = -\A\rr(t)+\xchi(t), 
\end{equation}
where $\A=\G^{-1}\U$ with $\G$ being
\begin{equation}
\label{matrixG}
\G = \gamma\left( \begin{array}{cc}
1 & -\kappa \\
\kappa & 1 \\
\end{array}\right),
\end{equation}
where $\gamma$ is the friction coefficient and  $\kappa=qB/\gamma$ is the diffusive Hall parameter which quantifies the strength of the Lorentz force relative to the frictional force. Here $\xchi(t)$ is Gaussian nonwhite noise~\cite{chun2018emergence} with zero mean and time correlation as
\begin{equation}
\label{timecorrelation}
\langle\xchi(t)\xchi^{\top}(t')\rangle= \G^{-1}\CC\delta_{+}(t-t')+ \CC(\G^{-1})^\top\delta_{-}(t-t'),
\end{equation}
where the notations $\delta_{\pm}(s')$ with $s'=t-t'$ indicate the modified Dirac delta functions which are zero for $s'\neq 0$ while $\int_0^\infty \dif s'\delta_+(s')=\int_{-\infty}^0\dif s'\delta_-(s')=1$ and $\int_0^\infty \dif s'\delta_-(s')=\int_{-\infty}^0\dif s'\delta_+(s')=0$. The matrix $\CC$ is given as 
\begin{equation}
\label{matrixC}
\CC = \frac{1}{2(1+\kappa^2)}\left( \begin{array}{cc}
2T_c+\kappa^2(T_c+T_h) & \kappa(T_h-T_c) \\
\kappa(T_h-T_c) & 2T_h+\kappa^2(T_c+T_h) \\
\end{array}\right).
\end{equation}

Note that the cross-correlation components of the correlation matrix in Eq.~\eqref{timecorrelation} have their origin in the broken time-reversal symmetry of the magnetic field~\cite{chun2018emergence,vuijk2019anomalous}.

\subsection*{Steady-State Properties}
\begin{figure}[t]
\centering
\includegraphics[width=1\linewidth]{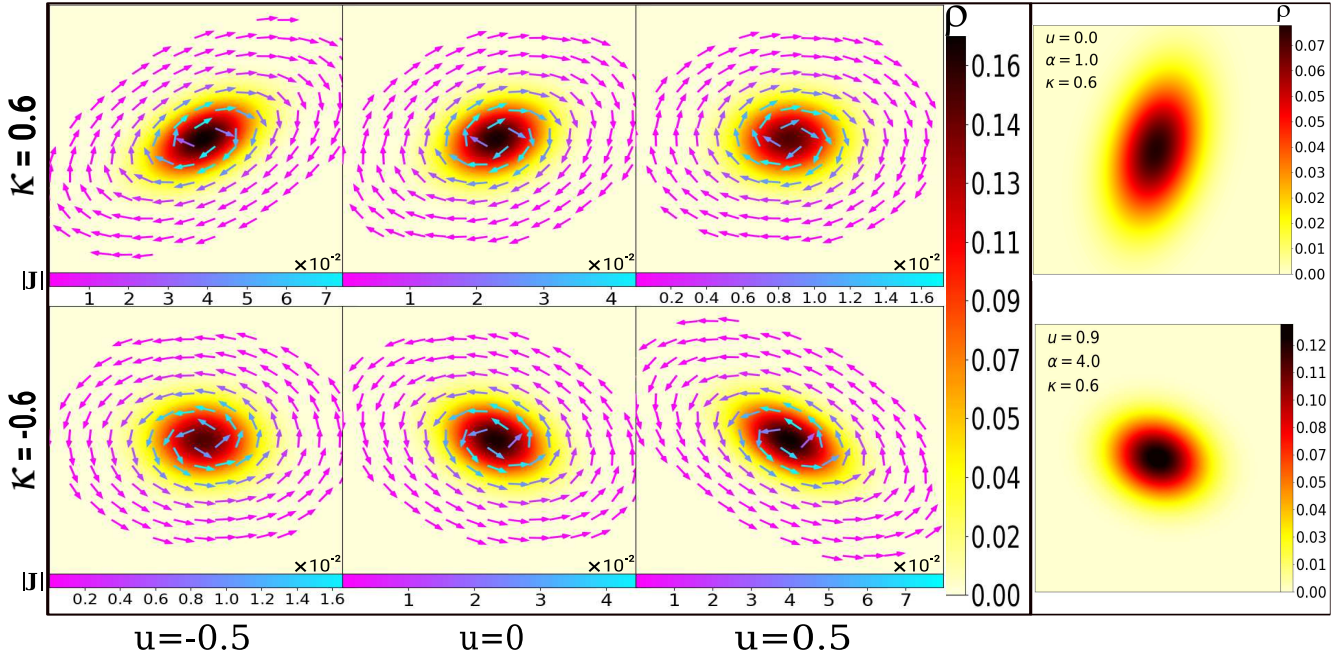}
\caption{The left figure shows a panel of the stationary-state probability density of the particle's position and fluxes in the system for different values of the diffusive Hall parameter $\kappa$ and the potential coupling parameter $u$. The results are shown from Eq.~\eqref{PDF} for the probability density and substituting that equation into Eq.~\eqref{flux} for the fluxes with $T_h=4T_c=4.0$ and $\alpha=4.0$. The figures correspond to $u=-0.5, 0.0, 0.5$ from  left to right for $\kappa=-0.6$ (top) and  $\kappa=0.6$ (bottom). The Brownian magneto-gyrator operates as a microengine even in the absence of the parameter $u$ due to the exerted torque by the particle on the potential. In this case, the direction of the gyration can be reversed by reversing the direction of the magnetic field resulting in a rotation in the probability density, shown in the middle figures. If $u\neq 0$ this can be done by additionally changing the sign of the potential coupling parameter or  tuning further the applied magnetic field. The direction of the fluxes are shown by arrows and the magnitude is color coded.
The right panel shows that the microengine can be stalled, which means that $\J=\0$, either
 if $\alpha=1$ and $u=0$ or $\kappa=2u/(\alpha-1)$ and $u\neq 0$ since the particle does not exert any torque. These cases are shown at the top and bottom, respectively. Note that the particle does not perform any gyrating motion if its motion is not correlated, namely when $u$ and $\kappa$ are zero. }
\label{fig:PDF}
\end{figure}

In order to investigate the steady-state properties of the system, we use the generalized Fokker-Planck equation corresponding to the overdamped Langevin equation~\eqref{langevinequaion:overdamped}, which can be written as

\begin{equation}
\label{FPE}
	\frac{\partial \rho(\rr, t)}{\partial t} = -\nabla\cdot\J(\rr, t),
\end{equation}
where $\rho(\rr, t)$ is the probability density of finding the particle at position $\rr$ at time $t$ and $\J(\rr, t)$ is the probability flux, given as
\begin{equation}
\label{flux}
\J(\rr, t)= -\D\nabla \rho(\rr, t)- (\A\cdot\rr)\rho(\rr, t),
\end{equation}
where the matrix $\D$ can be derived from Eq.~\eqref{langevinequaion:overdamped}~\cite{chun2018emergence} or alternatively by using a first-priciple approach ~\cite{abdoli2020correlations}, which is given as
\begin{equation}
\D
= \frac{1}{\gamma} 
\left( \begin{array}{cc}
\frac{T_c + \kappa^2 T_h}{\left(1+\kappa^2\right)^2} & \frac{\kappa(T_h - T_c)}{\left(1+\kappa^2\right)^2}+\frac{\kappa(T_c+T_h)}{2(1+\kappa^2)} \\
\frac{\kappa(T_h - T_c)}{\left(1+\kappa^2\right)^2}- \frac{\kappa(T_c+T_h)}{2(1+\kappa^2)} & \frac{T_h + \kappa^2 T_c}{\left(1+\kappa^2\right)^2} \\
\end{array}\right).
\label{diffusionmatrix}
\end{equation}
Surprisingly, despite the overdamped motion of the particles, the dynamics preclude a purely diffusive description as can be seen in the unusual structure of $\D$ above. In contrast to typical diffusion tensors, $\D$ has antisymmetric elements. These give rise to additional rotational fluxes in the system, which are perpendicular to the typical diffusive fluxes~\cite{vuijk2019anomalous, vuijk2019effect, abdoli2020nondiffusive, abdoli2020stationary, abdoli2021stochastic}. One of the known features of Lorentz forces is the reduction of the diffusion by a factor of $1/(1+\kappa^2)$, as can be seen from Eq.~\eqref{diffusionmatrix} for $T_c=T_h=T$.  Note that under Lorentz force, the diffusion is spatially correlated as can be seen from the symmetric, off-diagonal terms which are non-zero only for $T_h \neq T_c$ and $\kappa \neq 0$. 


Throughout this work we set the Boltzmann constant $k_B$ to unity. The stationary-state solution to Eq.~\eqref{FPE} is a Gaussian distribution, which is given as
\begin{equation}
\label{PDF}
\rho(\rr) = \frac{1}{Z}e^{-\frac{1}{2}(\rr^\top\cdot\M^{-1}\cdot\rr)},
\end{equation}
where $\M$ is the stationary-state covariance matrix and $Z$ is the normalization factor implying the probability density function with total probability of one, given as
\begin{equation}
\label{normalization}
Z=\frac{2\pi\sqrt{[u(T_h-T_c)+\kappa(\alpha T_c+T_h)]^2+(1+\alpha)^2T_cT_h}}{k(1+\alpha)\sqrt{(1+\kappa^2)(\alpha-u^2)}}, \hspace{1cm} \text{and} \hspace{1cm}
\M^{-1}=\left( {\begin{array}{*{20}c}
   \mu_1 & \mu_3  \\
   \mu_3 & \mu_2  \\    
 \end{array} } \right),
\end{equation}
where 
\begin{eqnarray}
     \label{mu1}
	\mu_1 &=& k(1+\alpha)\frac{(1+\kappa^2)T_h-u^2(T_h-T_c)+\alpha(\kappa^2 T_c+T_h)]}{[u(T_h-T_c)+\kappa(\alpha T_c+T_h)]^2+(1+\alpha)^2T_cT_h},\\\label{mu2}	
	\mu_2 &=& k(1+\alpha)\frac{\alpha^2(1+\kappa^2)T_c+u^2(T_h-T_c)+\alpha(T_c+\kappa^2T_h)}{[u(T_h-T_c)+\kappa(\alpha T_c+T_h)]^2+(1+\alpha)^2T_cT_h},\\
	\label{mu3}
	\mu_3 &=& -k(1+\alpha)\frac{u(\alpha T_c+T_h)(1+\kappa^2)-\kappa(\alpha-u^2)(T_h-T_c)}{[u(T_h-T_c)+\kappa(\alpha T_c+T_h)]^2+(1+\alpha)^2T_cT_h}.
\end{eqnarray} 

Note that in the case of an isotropic potential with $u=0$ and $\alpha=1$ Eqs.~\eqref{PDF} to \eqref{mu3} reduce to the results reported in Ref.~\cite{abdoli2020correlations} .  The probability fluxes in the stationary state of the system can be determined by substituting Eq.~\eqref{PDF} into Eq.~\eqref{flux}.

In the left panel of Fig.~\ref{fig:PDF} we show the stationary-state probability density of the position of the particle and the fluxes in the system for different values of the diffusive Hall parameter $\kappa$ and the potential coupling parameter $u$.  The fluxes are shown on top of the probability density. The Brownian magneto-gyrator does not essentially require a positional coupling due to the potential, as shown in the middle figures for $u=0$. However, the generic scenario in the presence of the potential coupling parameter, corresponding to $u\neq 0$, is also investigated that is shown in the left and right figures. The direction of the gyration of the microengine can be reversed by tuning the strength and direction of the applied magnetic field. In doing so, one rotates the probability density as well.  


\subsection*{Exerted Torque}


To obtain a simple scalar quantifier for the strength of the gyrating current field, which further emphasizes the tunability of the magneto-gyrator, we investigate the average torque on the potential. The average exerted torque by the particle on the potential $V$, denoted by $\langle\tau\rangle$, can be written as

\begin{equation}
\label{torque:definition}
\langle\tau\rangle = \int \rho(\rr) (\rr\times\F_c) \dif\rr,
\end{equation}
which is exactly equal to the opposite torque which the particle exerts via the friction forces on the thermal environment.
The substitution of Eq.~\eqref{PDF} into Eq.~\eqref{torque:definition} gives the average exerted torque on the potential by the particle as 
\begin{equation}
\label{torque}
\langle\tau\rangle = \frac{(T_h-T_c)[2u+(1-\alpha)\kappa]}{(1+\alpha)(1+\kappa^2)}.
\end{equation}
\begin{figure}[t]
\centering
\includegraphics[width=0.95  \linewidth]{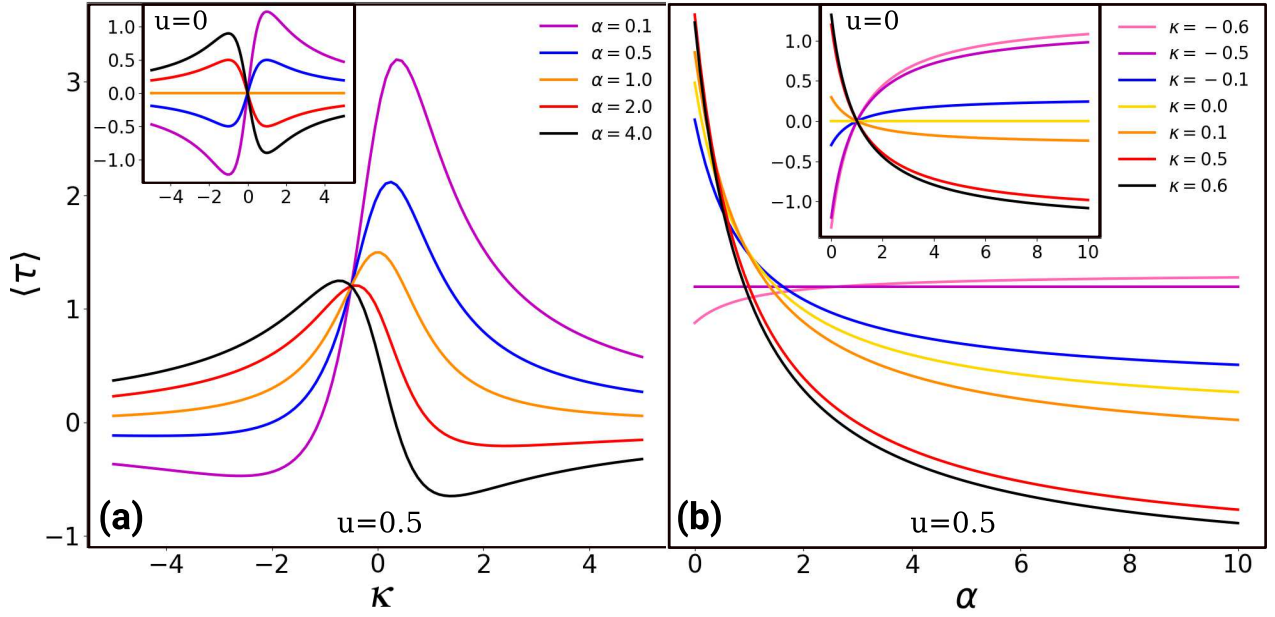}
\caption{ Average exerted torque by the microengine as a function of (a) diffusive Hall parameter $\kappa$ for different values of $\alpha$ and (b) the parameter $\alpha$ for different values of $\kappa$ with $T_h=4T_c=4.0$. The main figures show the results for $u=0.5$ while the insets depict those for $u=0$. As shown in (a), the average torque from Eq.~\eqref{torque} shows a maximum at an optimal $\kappa$, which can be reversed by tuning the strength and direction of the applied magnetic field. (b) In the generic scenario when $u\neq 0$, with increasing $\alpha$ the average torque increases if $\kappa<-0.5$ and decreases if $\kappa>-0.5$ till the exerted torque changes the direction resulting in the reversal of the gyration. It is independent of $\alpha$ if $\kappa=-0.5$. For $u=0$ with increasing $\alpha$, $\langle\tau\rangle$ increases if $\kappa<0$, decreases if $\kappa>0$ till the torque changes the sign, and is zero when $\kappa=0$. In addition, there is no torque if $\alpha=1$ and $\kappa\neq 0$, which correspond to the stalled microengine. }
\label{fig:torque}
\end{figure}


Note that Eq.~\eqref{torque} implies that even if $u=0$ there still exists a systematic torque giving rise to a nonequilibrium stationary state carrying fluxes, corresponding to an operating magneto-gyrator. This is due to the applied magnetic field ($\kappa\neq 0$) and the asymmetry of the potential about the origin ($\alpha\neq 1$). In this case, the torque exerted by the microengine on the potential can be reversed by reversing the direction of the applied magnetic field. This results in the reversal of the gyration, which is shown in the middle figures of the left panel of Fig.~\ref{fig:PDF}. In the presence of the potential coupling parameter, this can be done by additionally changing the sign of the potential coupling parameter or tuning further the applied magnetic field, shown in the left and right figures of Fig.~\ref{fig:PDF}. 
The average exerted torque is zero and consequently the microengine is stalled in two cases: (i) in the absence of the potential coupling parameter if $\alpha=1$  and (ii) if the potential is anisotropic with a nonzero $u$, but $\kappa=2u/(\alpha-1)$. The absence of fluxes in these cases are due to the cessation of the average torque which correspond to stalled microengines. The results for the  two cases are shown in the right panel of Fig.~\ref{fig:PDF}.

In Fig.~\ref{fig:torque} (a), we show that there exists a maximum torque exerted by the micro engine at an optimal magnetic field. The average torque can be reversed by reversing the direction of the magnetic field when $u=0$, which is shown in the inset. In the presence of the potential coupling parameter, shown in the main figure, this can be done by tuning further the applied magnetic field. 
Figure~\ref{fig:torque} (b) shows how the exerted torque by the microengine varies by tuning the stiffness of the potential for different values of the diffusive Hall parameter.
The main figure represents the results for the Brownian magneto-gyrator without the potential coupling parameter while those for the generic scenario, i.e., in the presence of $u$, are shown in the inset. From Eq.~\eqref{torque} it is clear that the average torque is independent of the parameter $\alpha$ if $u+\kappa=0$, which is shown in the main panel. The exerted torque decreases (increases) with increasing $\alpha$ if $\kappa>-u$ ($\kappa<-u$) till it changes the direction, corresponding to the change of the direction of the gyration. A similar explanation holds when $u=0$. However, there is no torque if $\kappa=0$ or $\alpha=1$ and consequently the microengine is stalled.  
   

\subsection*{Mechanical Power}

Now we consider the microengine in the presence of a load in order to determine the delivered mechanical power by the engine. To do this, we apply a linear external nonconservative force ($\nabla\times\F_{nc}\neq\0$) of the form $\F_{nc} = \epsilon (y, -x)$ with a parameter $\epsilon$, yielding a torque in the $z$ direction, whose sign is chosen such that the resulting torque in the $z$ direction is opposed to $\langle\tau\rangle$ in Eq.~\eqref{torque}. The goal is to calculate the average power of the work done by this force. 
The average mechanical power $P=\langle\F_{nc}.\dot{\rr}\rangle$ in the stationary state can be rewritten as $P=-\epsilon\langle xv_y-yv_x\rangle$. Therefore, one needs to calculate  the stationary-state position-velocity correlation matrix, that is $\lim_{t\rightarrow\infty}\langle\rr(t)\dot{\rr}^\top(t)\rangle$. To do this, one uses the overdamped Langevin equation~\eqref{langevinequaion:overdamped} for a modified matrix $\U_l$ under loading as
 \begin{equation}
\label{matrixU:modified}
\U_l = k\left( \begin{array}{cc}
1 & u-\epsilon^\prime \\
u+\epsilon^\prime & \alpha \\
\end{array}\right). 
\end{equation}
where $\epsilon^\prime=\epsilon/k$ is a dimensionless parameter. Using the equation of motion in Eq.~\eqref{langevinequaion:overdamped} it can be shown that 
\begin{equation}
\label{correlationmatrix}
\langle\rr\dot{\rr}^\top\rangle = -\langle\rr\rr^\top\rangle\A_l^\top + \langle\rr\xchi^\top\rangle,
\end{equation}
where $\A_l=\G^{-1}\U_l$. Note that by applying a prescribed external vortex
flow field such as a rotating bucket to an underdamped Brownian particle one can induce similar terms to the nonconservative force in the overdamped limit~\cite{liebchen2019optimal}. Moreover, it is known that in experiments using optical tweezers, optical scattering forces actually generate a nonconservative component to the overall force  exerted by the trap and do not simply furnish a simple potential~\cite{mangeat2019role}. Here we take a linear nonconservative force for the convenience of analytical calculations.
As we explain in the Methods, the  correlation matrices can be calculated and finally the average mechanical power can be written as 
\begin{equation}
\label{mechanicalpower}
P = \frac{h}{1-\lambda \frac{\epsilon}{\epsilon_s}}\frac{\epsilon}{\epsilon_{s}}\left(1-\frac{\epsilon}{\epsilon_s}\right),
\end{equation}
where $h=2\epsilon_s^2(T_c+T_h)/(k\gamma(1+\alpha))$, $\lambda=2\kappa\epsilon_s/(k(1+\alpha))$, and $\epsilon_s$ is the stall parameter which quantifies the maximum strength of the nonconservative force that one can apply to the engine without driving it in the inverse direction. Hence, the system operates as a heat engine, delivering mechanical work for $0 < \frac{\epsilon}{\epsilon_s} < 1$. The stall parameter is given as
\begin{equation}
\label{stallparameter}
\epsilon_s=-k\frac{\left (u-\frac{\kappa}{2}(\alpha-1)\right)\eta_c}{(2-\eta_c)(1+\kappa^2)},
\end{equation}
where $\eta_c=1-T_c/T_h$ is the Carnot efficiency. Having demonstrated in details the tunability of a Brownian magneto-gyrator in the presence of a rotated anisotropic potential, we focus from this point onwards on pure magneto-gyration, setting $u=0$.
\begin{figure}[t]
\centering
\includegraphics[width=0.8\linewidth]{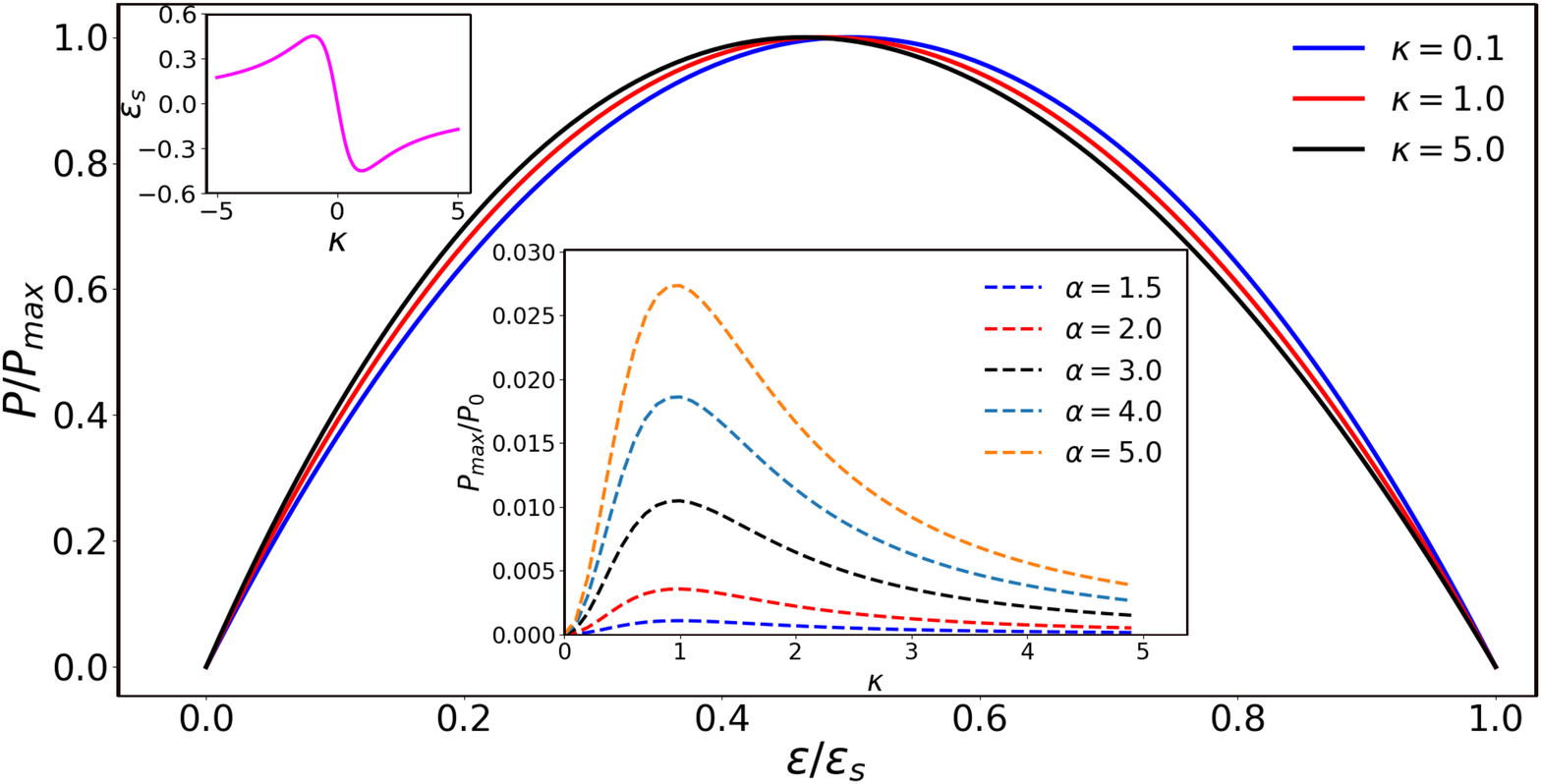}
\caption{The power of a Brownian magneto-gyrator with $T_h=4T_c=4.0$ and $u=0.0$. The main panel shows the scaled power $P/P_{max}$ as a function of $\epsilon/\epsilon_s$. The mechanical power vanishes when $\epsilon$ is zero or equal to the stall value. In addition, it has a maximum at $\epsilon/\epsilon_s=(1-\sqrt{1-\lambda})/\lambda$ where $\lambda=2\kappa\epsilon_s/(k(1+\alpha))$ which can be shifted by tuning the magnetic field. For the upper inset we use Eq.~\eqref{stallparameter} to plot the stall parameter with respect to the diffusive Hall parameter with $\alpha=4.0$. The stall parameter approaches zero with increasing magnitude of the applied magnetic field. In the lower inset, we show how the maximum power varies with respect to the diffusive Hall parameter for different values of $\alpha$.  $P_{max}$ is plotted in units of $P_0=(T_c+T_h)/\gamma$. There is an optimal value of $\kappa$ for which the maximum power takes its highest value.
}

\label{fig:work}
\end{figure}

The main panel of Fig.~\ref{fig:work} shows the scaled power $P/P_{max}$ delivered by the magneto-gyrator as a function of scaled loading $\epsilon/\epsilon_s$. As expected, the mechanical power vanishes when the engine is unloaded, i.e., $\epsilon=0$ or when the engine is stalled at $\epsilon=\epsilon_s$. The engine operates at maximum power at an intermediate loading $\epsilon/\epsilon_s=(1-\sqrt{1-\lambda})/\lambda$ which can be shifted by tuning the applied magnetic field. As can be seen in the inset to Fig.~\ref{fig:work}, there exists an optimal value of $\kappa$ for which the microengine delivers the maximum mechanical power. Qualitatively, this can be understood as follows. In the limit of vanishing magnetic field, there is no spatial correlation and hence no gyration. In the opposite case of very large magnetic field, the correlations vanish due to the large reduction in the diffusion coefficient of the particle.

\subsection*{Effect of Inertia on Power and Efficiency}
\begin{figure}[t]
\centering
\includegraphics[width=0.8\linewidth]{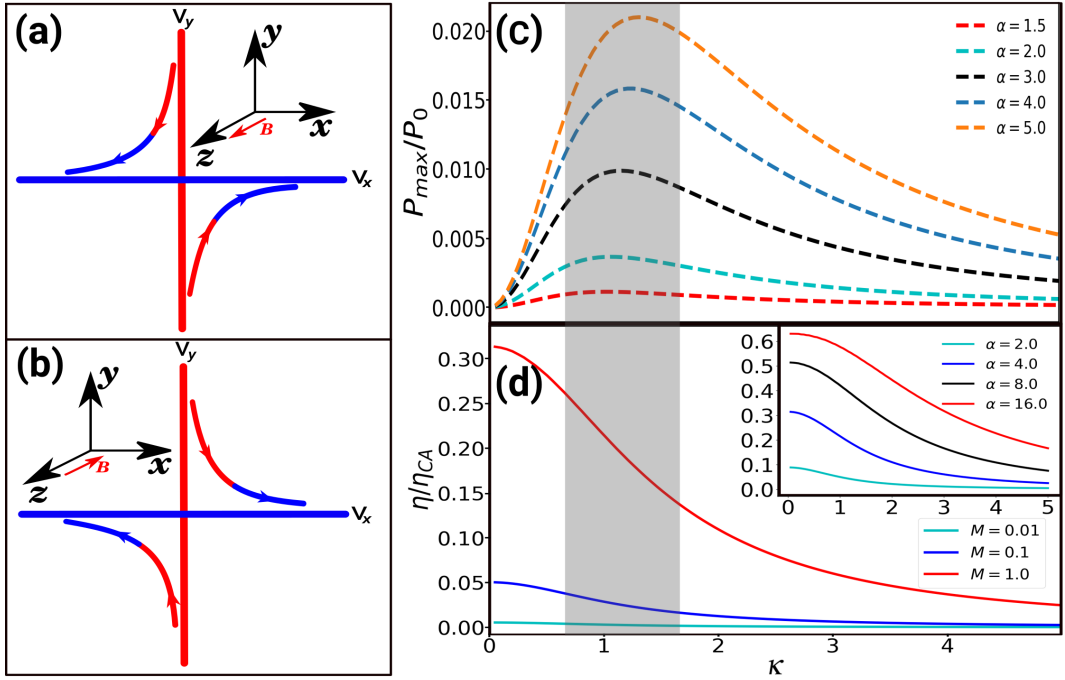}
\caption{(a) and (b) show the schematics of the heat transfer in the velocity space due to an external magnetic field in the $z$ and $-z$ directions, respectively. The velocity component which is coupled to the hot heat bath picks up higher kinetic energy, which due to the Lorentz force, gets transferred to the velocity component which is coupled to the cold heat bath on the time scale $m/\gamma$. 
This results in a heat transfer from the hot to the cold bath, shown in red and blue, respectively, in the velocity space. In (c), we represent the scaled maximum power with respect to the diffusive Hall parameter for a particle with $M=1.0$, $u=0.0$, and $T_h=4T_c=4.0$ for different values of the parameter $\alpha$ where $M=km/\gamma^2$ and $P_0=(T_c + T_h)/\gamma$. 
The maximum power decreases with increasing inertia. Inertia also changes the optimal diffusive Hall parameter at which the maximum power can be reached. Note that the plots reduce to the overdamped case in the small-mass limit, shown in the lower inset of Fig.~\ref{fig:work}. In  (d) the scaled efficiency $\eta/\eta_{CA}$ is shown in terms of the parameter $\kappa$ for $\alpha=4.0$, $T_h=4T_c=4.0$, and different values of $M$. The inset of (d) depicts the scaled efficiency for $M=1.0$ and different values of $\alpha$. The shaded regions in (c) and (d) mark the $\kappa$ region in which the magneto-gyrator is operating close to maximum power and the corresponding efficiency.
}
\label{fig:heatdissipation}
\end{figure}

While operating as a microengine, the magneto-gyrator dissipates heat at a steady rate while delivering output power. The rate of heat loss from the hot to the cold bath determines the efficiency $\eta$ of the magneto-gyrator. Surprisingly, as we show below, the heat dissipation in a magneto-gyrator cannot be obtained in the overdamped limit, which therefore always provides only an approximative description of the Brownian magneto-gyrator. In fact, the rate of heat loss in a magneto-gyrator requires explicit knowledge of the particle's mass and therefore has no counterpart in the overdamped approximation, where inertial effects are negligible. Qualitatively, the origin of heat loss can be understood as follows: in a magneto-gyrator the coordinate coupled to the hot thermostat picks up higher kinetic energy which, due to the Lorentz force, gets transferred to the coordinate coupled to cold thermostat on the time scale $m/\gamma$ (see Fig.~\ref{fig:heatdissipation} (a) and (b)). The steady-state is thus characterised by correlations between different velocity components~\cite{abdoli2020correlations}.

To calculate the efficiency and to quantify the effect of inertia on the extracted power of a magneto-gyrator, we model the dynamics of the particle via the underdamped Langevin equations, 
\begin{eqnarray}
     \label{underdampedx}
	m\dot{v}_x &=& -\gamma v_x+\gamma\kappa v_y-kx+\epsilon y+\xi_x(t) ,\\\label{underdampedy}	
	m\dot{v}_y &=& -\gamma v_y-\gamma\kappa v_x-k\alpha y - \epsilon x+\xi_y(t) ,
\end{eqnarray}
where $v_x=\dot{x}$ and $v_y=\dot{y}$ are the velocities of the particle in $x$ and $y$ directions, respectively. The particle is confined via the potential $k(x^2+\alpha y^2)/2$ and performs work against the external force $\epsilon(y,-x)$. For the sake of simplicity, we have chosen the potential coupling parameter $u$ to be zero. The absorbed heat by the particle from the cold and the hot baths, connected to the $i=x, y$ degrees of freedom,\, $\dbar Q_c$ and  \,$\dbar Q_h$, respectively, can be written as  
\begin{equation}
\label{absorbedheat}
\dbar Q_i=v_i\circ\left(-\gamma v_i \dif t + \dif W_i\right),
\end{equation}
where $\dif W=\int_t^{t+\dif t}\xi_i(t')\dif t'$ with zero mean and $\langle\left(\dif W_i\right)^2\rangle=2\gamma T_i\dif t$ and $\circ$ indicates the product in the Stratonovich sense. Using Eq.~\eqref{underdampedx} and Eq.~\eqref{underdampedy} the heat in Eq.~\eqref{absorbedheat} can be rewritten as
 \begin{eqnarray}
     \label{heatx}
	\dbar Q_c &=& \dif\left(\frac{1}{2}mv_x^2+\frac{k}{2}x^2-\frac{\epsilon}{2}xy\right)-\gamma\kappa v_xv_y\dif t-\frac{\epsilon}{2}(yv_x-xv_y)\dif t ,\\\label{heaty}	
	\dbar Q_h &=& \dif\left(\frac{1}{2}mv_y^2+\frac{k\alpha}{2} y^2+\frac{\epsilon}{2}xy\right)+ \gamma\kappa v_xv_y\dif t-\frac{\epsilon}{2}(yv_x-xv_y)\dif t.
\end{eqnarray}
The total derivatives in Eq.~\eqref{heatx} and Eq.~\eqref{heaty} have no contribution to the steady-state average. Therefore, the average mechanical power can be calculated (see the Supplemental Materials), which in this case is given as
\begin{equation}
\label{ssheat_inertia}
P = \frac{4k(T_c+T_h)\frac{\epsilon}{\epsilon_s}\left(1-\frac{\epsilon}{\epsilon_s}\right)}{\gamma\left[G_1 - 4M(\frac{\epsilon}{\epsilon_s}+ G_2)^2\right]},
\end{equation}
where the mass-independent stall parameter $\epsilon_s$ is given by Eq.~\eqref{stallparameter} with $u=0$, and $G_1=k^2(1-\alpha)^2M/(\epsilon_s^2(1+\kappa^2))+k^2(2M(1+\alpha)+\kappa^2)/(\epsilon_s^2 M)$ and $G_2=k\kappa/(2\epsilon_s M)$ are dimensionless. Here $M=km/\gamma^2$ is a dimensionless parameter. Note that Eq.~\eqref{ssheat_inertia} reduces to Eq.~\eqref{mechanicalpower} in the small-mass limit. In Fig.~\ref{fig:heatdissipation} (c), we show the maximum power of a magneto-gyrator made of a charged Brownian particle with $M=1.0$, $u=0.0$, $T_h=4T_c=4.0$ with respect to the diffusive Hall parameter for  different values of the parameter $\alpha$. 
Comparing to the overdamped case in Fig.~\ref{fig:work}, we find that inertia shifts the optimum magnetic field at which the maximum power is delivered. Moreover, inertia reduces the maximal power that a magneto-gyrator can deliver.

The steady-state average of the heat flux from the hot bath to the cold one can be read from Eqs.~\eqref{heatx} and \eqref{heaty} as $\langle\dot{Q}\rangle_{ss}=\gamma\kappa\langle v_xv_y\rangle$, 
where the dot over $Q$ indicates the time derivative. Since the velocity correlation due to the magnetic field depends on the mass of the particle, the heat loss can be written as $\langle\dot{Q}\rangle=f(\kappa,M,\epsilon)/M$. The function $f$ and the details are given in the Supplemental Materials. In the limit of zero mass, the function $f$ remains finite giving rise to a divergent heat loss. Therefore, even in the overdamped regime the knowledge of particle's mass is needed for the calculation of the heat loss and consequently the efficiency $\eta=P/\langle\dot{Q}\rangle$ of the microengine. 

Carnot efficiency can be obtained for an infinitely slow transformation, which for the magneto-gyrator corresponds to a stalled engine with large $M$. Due to this, a more useful notion is that of efficiency at maximum power~\cite{benenti2011thermodynamic}. The efficiency at maximum power is bounded from above as $\eta_{CA} = 1 - \sqrt{T_c/T_h}$, commonly referred to as the Curzon-Ahlborn upper bound~\cite{curzon1975efficiency}. In Fig.~\ref{fig:heatdissipation} (d) the scaled efficiency $\eta/\eta_{CA}$ is shown in terms of the parameter $\kappa$ for $\alpha=4.0$, $T_h=4T_c=4.0$, and different values of $M$. As expected, the efficiency at maximum power increases with increasing $M$ due to the reduction in heat loss in the velocity space. One could operate the magneto-gyrator as following: For a given $M$ and $\alpha$, tune the magnetic field so that the magneto-gyrator operates at the maximum power as shown in the shaded region in Fig.~\ref{fig:heatdissipation} (c) with a corresponding efficiency at maximum power in Fig.~\ref{fig:heatdissipation} (d). Note that the dimensionless mass $M=1.0$ is achievable for a particle with a radius $R\sim 10^{-5}m$ and mass $m\sim 10^{-11} kg$ (such as spherical PMMA particles) in a potential with an optical trapping stiffness $k\sim 10^{-6}N/m$, as used in Ref.\cite{argun2017experimental}, in a solvent of a viscosity $\nu\sim 10^{-5} N.s/m^2$ such as a dusty plasma. This implies that for the parameters $\alpha = 4$, $u = 0$ and $M = 1$, magneto-gyrator will deliver maximum possible power at $\kappa \approx 1$ with an efficiency $\approx 0.2\eta_{CA}$ for a experimentally realisable temperature ratio $T_h/T_c = 4.0$~\cite{argun2017experimental}.

\section*{Discussion}

In this paper, we proposed a Brownian magneto-gyrator as a tunable microengine. The microengine can be steered via an external magnetic field, which allows tuning of the key properties, such as the direction of gyration, the torque, and the output power by varying the strength and the direction of the applied magnetic field. The working principle of the proposed magneto-gyrator is the steady state correlations between the spatial degrees of freedom. These correlations are induced by the Lorentz force and do not require special alignment of the confining potential with respect to the temperature axes, as in the previous studies~\cite{filliger2007brownian, dotsenko2013two,mancois2018two, cerasoli2018asymmetry, murashita2016overdamped, holubec2017thermal, nascimento2020memory, nascimento2020stationary, chang2021autonomous}. In fact, Lorentz force-induced correlations exist even for a freely diffusing Brownian particle subjected to two different thermostats~\cite{abdoli2020correlations}. The correlations, however, vanish for large magnetic fields; the diffusion becomes smaller with increasing magnetic field. This implies that there exists optimal magnetic field to operate the magneto-gyrator as a microengine, as reflected by the existence of a maximum in the output power as a function of the diffusive Hall parameter.

An intriguing feature of a microengine is the heat loss in the system. For a magneto-gyrator, as we showed, there is a heat loss not only in the position space but also in the velocity space. The heat loss in the velocity space depends on the particle's mass and therefore has no counterpart in the overdamped approximation, where inertial effects are negligible compared to the frictional forces. As a consequence, an explicit knowledge of the mass of the particle is needed for the calculation of the efficiency of the microengine. By going beyond the overdamped approximation, we investigated the effect of inertia on the power and efficiency of the microengine.

Finally, we consider the possibility of an experimental realisation of the proposed magneto-gyrator. A possible experimental set up is to trap the particle using optical tweezers subjected to a fluctuating electric field either in a radio-frequency plasma sheath with a vertical magnetic field~\cite{carstensen2009effect, piel2017plasma} or in a rotating frame of reference. In the latter case, a well-controlled rotation of the reference frame induces a Coriolis force which acts the same as the Lorentz force due to an external magnetic field~\cite{kahlert2012magnetizing, hartmann2013magnetoplasmons, hartmann2019self}. Based on these experimental studies, it seems that a magneto gyrator could be realised in dusty plasmas with an efficiency at maximum power of $\sim 0.2 \eta_{CA}$ while delivering maximum power at experimentally realisable magnetic fields. From a future perspective, we plan to go beyond the description of the magneto-gyrator in terms of its average properties. We plan to use stochastic thermodynamics to calculate the fluctuations in torque and work delivered by the magneto-gyrator. Furthermore, it would be interesting to generalize the notion of the magneto-gyrator to interacting particles. Finally, our analysis might be applicable to other systems which exhibit circular motion in an anisotropic fluctuation field including chiral colloidal microswimmers in parabolic potentials~\cite{jahanshahi2017brownian}, active Janus particles in a complex plasma~\cite{nosenko2020active}, particles dominated by the Magnus force~\cite{reichhardt2020dynamics} and even rotating skyrmions~\cite{zhang2018manipulation}.  

\section*{Methods}
\label{methods}
\subsection{Langevin Approach}
In this section, we take the small-mass approach to derive the overdamped Langevin equation describing a Brownian gyrator under the effect of Lorentz force. The two-dimensional motion of a charged, Brownian particle of mass $m$, trapped in an external potential $V(\rr)$, in the presence of an external magnetic field $B$ in the $z$ direction can be described by the following underdamped Langevin equation
\begin{equation}
\dot{\qq}(t) =  -\F\qq(t) + \tilde{\xxi}(t),
\label{langevinequation:q}
\end{equation}
where $\qq(t)=(x(t), y(t), v_x(t), v_y(t))^\top$ and $\tilde{\xxi}(t)=(0, 0, m^{-1}\xi_x(t), m^{-1}\xi_y(t))^\top$ is Gaussian white noise with zero mean and time correlation $\langle\tilde{\xxi}(t)\tilde{\xxi}^{\top}(t')\rangle= (2\gamma/m^2)\T\delta(t-t')$ where $\gamma$ is the constant friction coefficient. Here $\T=\diag(0, 0, T_c, T_h)$ is a diagonal  matrix and the matrix $\F$ is defined as
\begin{equation}
\label{methods:matrixF}
\F = \frac{1}{m}\left( \begin{array}{cc}
\0 & -m\1 \\
\U & \G \\
\end{array}\right), 
\end{equation}
where $\1$ is the identity matrix and
\begin{equation}
\label{methods:matrixG}
\G = \gamma\left( \begin{array}{cc}
1 & -\kappa \\
\kappa & 1 \\
\end{array}\right), \hspace{1cm}
\U = k\left( \begin{array}{cc}
1 & u \\
u & \alpha \\
\end{array}\right). 
\end{equation}
Taking the small mass approach, detailed in the Supplemental Material, yields the following overdamped Langevin equation
\begin{equation}
\label{methods:langevinequaion:overdamped}
\dot{\rr}(t) = -\A\rr(t)+\xchi(t), 
\end{equation}
where $\A=\G^{-1}\U$ and $\xchi(t)$ is nonwhite noise whose properties are given as
\begin{eqnarray}
     \label{methods:mean}
	\langle\xchi(t)\rangle &=& 0 ,\\\label{methods:timecorrelation}	
	\langle\xchi(t)\xchi^\top(t')\rangle &=& \G^{-1}\CC\delta_{+}(t-t')+ \CC(\G^{-1})^\top\delta_{-}(t-t') ,
\end{eqnarray} 
where $\delta_{\pm}(t-t')$ are different Dirac delta functions and the matrix $\CC$ is given by Eq.~\eqref{correlationmatrix}.

\subsection{Fokker-Planck Approach}
The Fokker-Planck equation for the probability density of finding the particle at position $\rr$ at time $t$, denoted by $\rho(\rr, t)$, corresponding the overdamped Langevin equation~\eqref{methods:langevinequaion:overdamped}, can be derived using different methods~\cite{gardiner2009stochastic, vuijk2019lorenz} which give
\begin{equation}
\label{methods:FPE}
	\frac{\partial \rho(\rr, t)}{\partial t} = -\nabla\cdot\J(\rr, t),
\end{equation}
where $\J(\rr, t)$ is the probability flux, given as
\begin{equation}
\label{methods:flux}
\J(\rr, t)= -\D\nabla \rho(\rr, t)- (\A\cdot\rr)\rho(\rr, t),
\end{equation}
where the matrix $\D$ is given as in Eq.~\eqref{diffusionmatrix}. Equation~\eqref{methods:FPE} is a linear Fokker-Planck equation whose Gaussian solution can be written as
\begin{equation}
\label{methods:gaussiansolution}
\rho(\rr, t)= \frac{1}{Z}e^{-\frac{1}{2}\left(\rr^\top\cdot\M_t^{-1}\cdot\rr\right)},
\end{equation}
where $Z=2\pi\sqrt{\Det(\M_t)}$ and the covariance matrix $\M_t$ satisfies the following Lyapunov equation
\begin{equation}
\label{methods:lyapunovequation}
\frac{\dif\M_t}{\dif t}=\A\M_t+\M_t\A^\top+2\D_s
\end{equation}
where $\D_s=(\D+\D^\top)/2$ is the usual (symmetric) diffusion matrix. Equation~\eqref{methods:lyapunovequation} can be easily solved for the stationary state by setting $\dif\M_t/\dif t$ to zero. This gives the stationary-state covariance matrix, denoted by $\M$, as
\begin{equation}
\label{methods:covariancematrix}
\M = \left( \begin{array}{cc}
\frac{\alpha^2(1+\kappa^2)T_c+u^2(T_h-T_c)+\alpha (T_c+\kappa^2T_h)}{k(1+\alpha)(\alpha-u^2)(1+\kappa^2)} & \frac{\kappa (\alpha-u^2)(T_h-T_c)-u(1+\kappa^2)(\alpha T_c+T_h)}{k(1+\alpha)(\alpha-u^2)(1+\kappa^2)}  \\
\frac{\kappa(\alpha-u^2)(T_h-T_c)-u(1+\kappa^2)(\alpha T_c+T_h)}{k(1+\alpha)(\alpha-u^2)(1+\kappa^2)} & \frac{(1+\kappa^2)T_h-u^2(T_h-T_c)+\alpha(\kappa^2T_c+T_h)}{k(1+\alpha)(\alpha-u^2)(1+\kappa^2)} \\
\end{array}\right), 
\end{equation}
where $\kappa=qB/\gamma$ is the diffusive Hall parameter which quantifies the strength of the Lorentz force relative to the frictional force. Plugging Eq.~\eqref{methods:covariancematrix} into Eq.~\eqref{methods:gaussiansolution} gives the stationary-state probability density as in Eq.~\eqref{PDF}.

\subsection{Average Mechanical Power}
In order to determine the average mechanical power $P=\langle\F_{nc}\cdot\dot{\rr}\rangle$ of the work done by  a linear nonconservative force $\F_{nc}=\epsilon (y, -x)$  yielding a torque in the $z$ direction, one needs to calculate the stationary-state moment covariance matrix.  Using the overdamped Langevin equation the matrix can be written as 
\begin{equation}
\label{methods:correlationmatrix}
\langle\rr\dot{\rr}^\top\rangle = -\langle\rr\rr^\top\rangle\A_l^\top + \langle\rr\xchi^\top\rangle,
\end{equation}
where $\A_l=\G^{-1}\U_l$ with $\U_l$ being the same as in Eq.~\eqref{matrixU:modified}. We eliminated the time dependence from the notations so as to indicate the stationary-state quantities. The position correlation and the position-noise correlation can be written as
\begin{eqnarray}
     \label{methods:positioncorrelation}
	\langle\rr\rr^\top\rangle &=& \lim_{t\rightarrow\infty}\int_0^t\dif t'\int_0^t\dif t''e^{-\A_l(t-t')}\langle\xchi(t')\xchi^\top(t'')\rangle e^{-\A_l^\top(t-t'')},\\\label{methods:positionnoisecorrelation}	
	\langle\rr\xchi^\top\rangle &=& \lim_{t\rightarrow\infty}\int_0^t\dif t' e^{-\A_l(t-t')}\langle\xchi(t')\xchi^\top(t)\rangle.
\end{eqnarray} 
The substitution of the properties of the Gaussian nonwhite noise from Eq.~\eqref{timecorrelation} into Eq.~\eqref{methods:positioncorrelation} and Eq.~\eqref{methods:positionnoisecorrelation} gives
\begin{eqnarray}
     \label{methods:2positioncorrelation}
	\langle\rr\rr^\top\rangle &=& \int_0^\infty\dif s e^{-\A_l s}[\G^{-1}\CC+\CC(\G^{-1})^\top]e^{-\A_l^\top s},\\\label{methods:2positionnoisecorrelation}	
	\langle\rr\xchi^\top\rangle &=& \CC(\G^{-1})^\top ,
\end{eqnarray} 
where Eq.~\eqref{methods:2positioncorrelation} is the solution to the following Lyapunov equation
\begin{equation}
\label{methods:lyapunov2}
\A_l\langle\rr\rr^\top\rangle+\langle\rr\rr^\top\rangle\A_l^\top=-\Xi,
\end{equation}
where $\Xi=\G^{-1}\CC+\CC(\G^{-1})^\top$. The solution to Eq.~\eqref{methods:lyapunov2} is given as
 \begin{equation}
\label{correlationmatrix}
\langle\rr\dot{\rr}^\top\rangle = \left( \begin{array}{cc}
0 & \frac{[u-\frac{\kappa}{2}(\alpha-1)](T_h-T_c)+\epsilon^\prime(1+\kappa^2)(T_c+T_h)}{\gamma(1+\kappa^2)(1+\alpha-2\epsilon^\prime\kappa)} \\
-\frac{[u-\frac{\kappa}{2}(\alpha-1)](T_h-T_c)+\epsilon^\prime(1+\kappa^2)(T_c+T_h)}{\gamma(1+\kappa^2)(1+\alpha-2\epsilon^\prime\kappa)} & 0 \\
\end{array}\right),
\end{equation}
and the average mechanical power reads
\begin{equation}
\label{methods:mechanicalpaower}
P=2k\epsilon^\prime\frac{[u-\frac{\kappa}{2}(\alpha-1)](T_h-T_c)+\epsilon^\prime(1+\kappa^2)(T_c+T_h)}{\gamma(1+\kappa^2)(1+\alpha-2\epsilon^\prime\kappa)},
\end{equation}
which can be rewritten as Eq.~\eqref{mechanicalpower}.

\providecommand{\noopsort}[1]{}\providecommand{\singleletter}[1]{#1}%


\end{document}